\documentclass[9pt,twocolumn,twoside,lineno]{pnas-new}
\pdfoutput=1

\templatetype{pnasinvited} % Choose template 
\title{Nanoporosity imaging by positronium lifetime tomography}

\author[1,2,*]{K.~Dulski}
\author[1,2]{E.~Beyene}
\author[1,2]{N.~Chug}
\author[3]{C.~Curceanu}
\author[1,2]{E.~Czerwi{\'n}ski}
\author[1,2]{M.~Das}
\author[4]{M.~Gorgol}
\author[4]{B.~Jasi{\'n}ska}
\author[1,2]{K.~Kacprzak}
\author[1,2]{\L{}.~Kap\l{}on}
\author[1,2]{G.~Korcyl}
\author[1,2]{T.~Kozik}
\author[1,2]{K.~Kubat}
\author[1,2]{D.~Kumar}
\author[5]{E.~Lisowski}
\author[5]{F.~Lisowski}
\author[1,2]{J.~M\k{e}drala-Sowa}
\author[1,2]{S.~Nied{\'z}wiecki}
\author[1,2]{P.~Pandey}
\author[1,2]{S.~Parzych}
\author[1,2]{E.~Perez del Rio}
\author[1,2]{M.~R\"{a}dler}
\author[1,2]{S.~Sharma}
\author[1,2]{M.~Skurzok}
\author[1,2]{K.~Tayefi}
\author[1,2]{P.~Tanty}
\author[1,2]{E.~{\L{}}.~St\k{e}pie{\'n}}
\author[1,2,*]{P.~Moskal}
\affil[1]{Faculty of Physics, Astronomy and Applied Computer Science, Jagiellonian University, S.~{\L}ojasiewicza 11, 30-348 Krakow, Poland}
\affil[2]{Centre for Theranostics, Jagiellonian University, Kopernika 40, 31-501 Krakow, Poland}
\affil[3]{INFN, Laboratori Nazionali di Frascati,  Via E. Fermi 40, 00044 Frascati, Italy}
\affil[4]{Institute of Physics, Maria Curie-Skłodowska University, Radziszewskiego 10, 20-031 Lublin, Poland}
\affil[5]{Cracow University of Technology, Warszawska 24, 31-864 Kraków, Poland}
\affil[*]{corresponding author, e-mail: \url{kamil.dulski@uj.edu.pl}}

\authorcontributions{}
\authordeclaration{}
\equalauthors{}
\correspondingauthor{\textsuperscript{*}To whom correspondence should be addressed. E-mail: kamil.dulski@uj.edu.pl or p.moskal@uj.edu.pl}
\doi{\url{www.doiToFillIn.com}}

\keywords{Positron, PALS, Nanoporosity imaging, Nanoporous materials} 

\begin{abstract}
Positron Annihilation Lifetime Spectroscopy (PALS) is a well-established non-destructive technique used for nanostructural characterization of porous materials. It is based on the annihilation of a positron and an electron. Mean positron lifetime in the material depends on the free voids size and molecular environment, allowing the study of porosity and structural transitions in the nanometer scale. We have developed a novel method enabling spatially resolved PALS, thus providing tomography of nanostructural characterization of an extended object. 
Correlating space (position) and structural (lifetime) information brings new insight in materials studies, especially in the characterization of the purity and pore distribution. For the first time, a porosity image using stationary positron sources for the simultaneous measurement of the porous polymers XAD4, silica aerogel powder IC3100, and  polyvinyl toluene scintillator PVT by the J-PET tomograph is demonstrated.

\end{abstract}

\dates{This manuscript was compiled on \today}

\begin{document}

\maketitle

\section*{INTRODUCTION}

\noindent The development of imaging techniques and the characterization of materials in three dimensions opens the possibility to study new structural relationships and the separation of changes in different layers of a sample. The most common techniques, such as Focused Ion Beam (FIB), X-ray tomography (X-ray CT), Laser Scanning Confocal Microscopy (LSCM) and Transmission Electron Microscopy (TEM) create image of the sample with a resolution up to micro- and nano-meters \cite{PorousImagingTechniques, PorousImagingTechniques2}. However, sample preparation in FIB, LSCM, and in TEM usually could result in destroying the studied object \cite{PorousImagingTechniques}. In addition, there are some limitations to the volume of the sample that can be imaged (up to $(10$  $\mu$m$)^3$ for FIB \cite{PorousImagingTechniques, PorousImagingTechniques2}) or requiring very thin samples (10 $\mu$m for TEM \cite{TEM}, 7 $\mu$m thick for LSCM \cite{PorousImagingTechniques}), the preparation of which is often very complicated and time-consuming. In contrast, X-ray tomography allows one to measure samples in a wide range of sizes \cite{PorousImagingTechniques2, mX-ray}. However, it allows only to characterize pores in the micrometer scale. Therefore, nanostructural information coming from the measurement of the mean positronium lifetime \cite{PALS_porosity1, PALS_porosity2} could become a complementary technique, allowing enhanced porosity imaging of the matter. Here we demonstrate a method of imaging of the nano-porosity in large objects (tens of cm) by a novel technique combining the Positron Annihilation Lifetime Spectroscopy (PALS) and Positron Emission Tomography (PET). The main novelty of the method lies in the determination of the annihilation location and the positron lifetime for each event separately, and to perform PALS analysis for each voxel of the reconstructed PET image.

\noindent PALS is a well-established technique that enables the study of porosity and phase transitions in matter such as porous silica, metals, compounds and semiconductors \cite{PALS_porosity1, PALS_porosity2, ChemPhys-LFvsSize-1972,ChemPhys-LFvsSize-1981,ConfSeries-LFvsSize-2013,ChemPhysLett-LFvsSize-1997, JPorousMatter-PALSsilica-2016, ApplPhys-PALSsilica-2001, DefDiffForum-PALSmetal-2011, AlloysComp-PALSmetal-2007, PhysSolidi-PALSsemicond-2013} and biological samples \cite{ACSNano-BiolPals-2021, SciRep-BiolPals-2023, EJNMMIPhys-BiolPals-2023, BAMS-BiolPals-2022}.
The annihilation of electron and positron may proceed via formation of a quasi-stable two-body system - positronium (Ps) with a mass of 1.022 MeV\slash c$^2$. Positronium, depending on its total spin (S), can be formed in two states: para-positronium (p-Ps, \textbf{S} = 0) or ortho-positronium (o-Ps, \textbf{S} = 1). Due to the conservation of charge conjugation symmetry singlet state p-Ps annihilates with emission of even number of photons, unlike triplet state o-Ps which annihilates with emission of odd number of photons \cite{RevModPhys-PositPhys-2023}. An additional factor distinguishing the two states of positronium is the average time after which they annihilate (mean lifetime). In a vacuum, para-positronium annihilates with mean lifetime of 0.125 ns \cite{PhysRevLett-pPsDecayRate-1994}, in contrast to ortho-positronium which annihilates after mean lifetime of 142 ns \cite{PhysLettB-oPsDecayRate-2009}. The mean lifetime of ortho-positronium decreases significantly in materials and depends on the material nano-structure \cite{PALS_porosity1, PALS_porosity2}. In PALS the lifetime of a~positron is usually extracted by using a specific isotope (e.g.$^{22}$Na) which, after $\beta ^+$ decay, enters the excited state of the daughter nuclide, which then deexcites with emission of gamma quantum, hereinafter referred to as a deexcitation photon. Time difference between the emission of annihilation and deexcitation photon can be used as an estimate of the lifetime of the positron that annihilated.

\noindent Mean ortho-positronium lifetime is a sensitive probe to the structure and chemical environment in which it is located. The relationship between the mean lifetime of a positron and the size of free volumes (pores) is the basis of PALS \cite{ChemPhys-LFvsSize-1972,ChemPhys-LFvsSize-1981,ConfSeries-LFvsSize-2013,ChemPhysLett-LFvsSize-1997}. The shortening of the mean ortho-positronium lifetime can be translated into the radius of free volume using the Tao-Eldrup model \cite{ChemPhys-LFvsSize-1972,ChemPhys-LFvsSize-1981} or its extensions \cite{ConfSeries-LFvsSize-2013,ChemPhysLett-LFvsSize-1997}, for inorganic matter measured in a vacuum. It was also found that in the presence of free radicals, the mean lifetime of o-Ps may be additionally shortened due to the o-Ps to p-Ps conversion processes \cite{PCCP-PALSOxygen-2020,CommPhys-PALSOxygen-2020}.

\noindent PET is used in medical diagnostics for imaging of the metabolism rate of radiopharmaceuticals in tissues of the living organism \cite{BAMS-PETImpact-2021, BAMS-PETRev-2021, PETCLin-TBJPET-2020, PETClin-PET-2020, EJNMMI-PET-2020, EJNMMI-TBPET-2020}. In PET, radiopharmaceuticals are labeled with $\beta ^+$ radionuclide, and photons emitted from the body due to the positron-electron annihilations are used to reconstruct an image of the annihilation density distribution which reflects the metabolic rate. Recently, the Jagiellonian PET group (J-PET), had reported the idea \cite{IEEE-JPET-2019} and simulation-based feasibility study of combining PALS and PET techniques in the so-called positronium imaging \cite{Nature-PositronMedicine-2019, PMB-Morpho-2019, EJNMMI-Morpho-2020, Science-Morpho-2021, Science-Morpho-2024}. Additionally, the development of imaging reconstruction techniques specially adapted to positronium imaging allows obtaining image resolution comparable to those obtained in standard PET \cite{Shopa2022, Qi2022, Huang2022, Huang2024, BAMS-PosImaReco-2023, BAMS-PosImaReco2-2023}.

\begin{figure}[H] 
    \centering
    \includegraphics[width=0.48\textwidth]{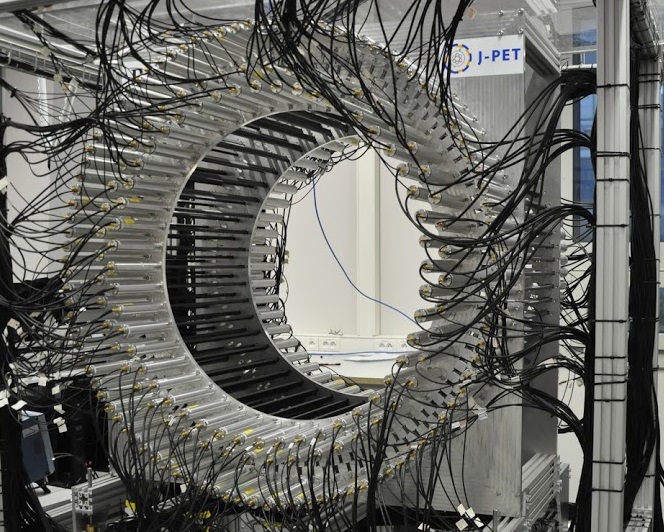}
    \caption{Photograph of the J-PET detector  constructed in the Jagiellonian University \cite{ActaB-JPET-2017}. 192 polymer scintillation strips with 50 cm length (black strips) are ordered in three concentric cylindrical layers with radius 42.5, 46.75 and 57.5 cm. Both ends of the scintillation strip is connected to the Hamamatsu R9800 vacuum photomultipliers (grey tubes) \cite{NIM-SingleModuleJPET-2014, NIM-TOF-JPET-2015}. Light signal created in the scintillators are converted to electric signals by photomultipliers and then transferred to the data acquisition system, digitized \cite{JINST-FPGA-2017} and acquired by the dedicated FPGA system \cite{IEEE-FPGA-2018}.}
    \label{fig:JPET}
\end{figure}

\noindent In this work, mean o-Ps lifetime images will be demonstrated for three samples of different porosity - the porous polymers XAD4, silica aerogel powder IC3100, and polyvinyl toluene scintillator PVT. It will be the first image of this type obtained for porous samples, which allows for the tomographic imaging of the free void size distributions in large extended objects. The designed positronium image reconstruction algorithm for the J-PET detector made it possible to obtain a resolution of several centimeters.

\section*{THE J-PET DETECTOR AND MEASUREMENT DETAILS}

The J-PET (Jagiellonian Positron Emission Tomography) detector (Fig.~\ref{fig:JPET}) was constructed from
%as an alternative to commercial PET scanners. Commonly used small inorganic scintillators were replaced with 
long EJ-230 plastic scintillator strips \cite{NIM-SingleModuleJPET-2014, NIM-TOF-JPET-2015, ActaB-JPET-2017}. Dimensions of each strip are equal to $1.9 \times 0.7 \times 50 \text{ cm} ^3$.
%, allowing not only to reduce the cost of the device, but also to introduce new solutions and features to medical diagnostics
Scintillator strips in the detector are arranged in three concentric cylindrical layers (with radii 42.5, 46.75 and 57.5 cm) \cite{NIM-SingleModuleJPET-2014, NIM-TOF-JPET-2015, ActaB-JPET-2017}. The long strips and large radii of the layers allows to measure large samples, even up to 40$\times$40$\times$40 cm$^3$. Light signals generated inside the scintillation strips are converted into electric signals by Hamamatsu R9800 vacuum photomultipliers coupled to both ends of the strip.

\begin{figure}[H] 
    \centering.
    \includegraphics[width=0.48\textwidth]{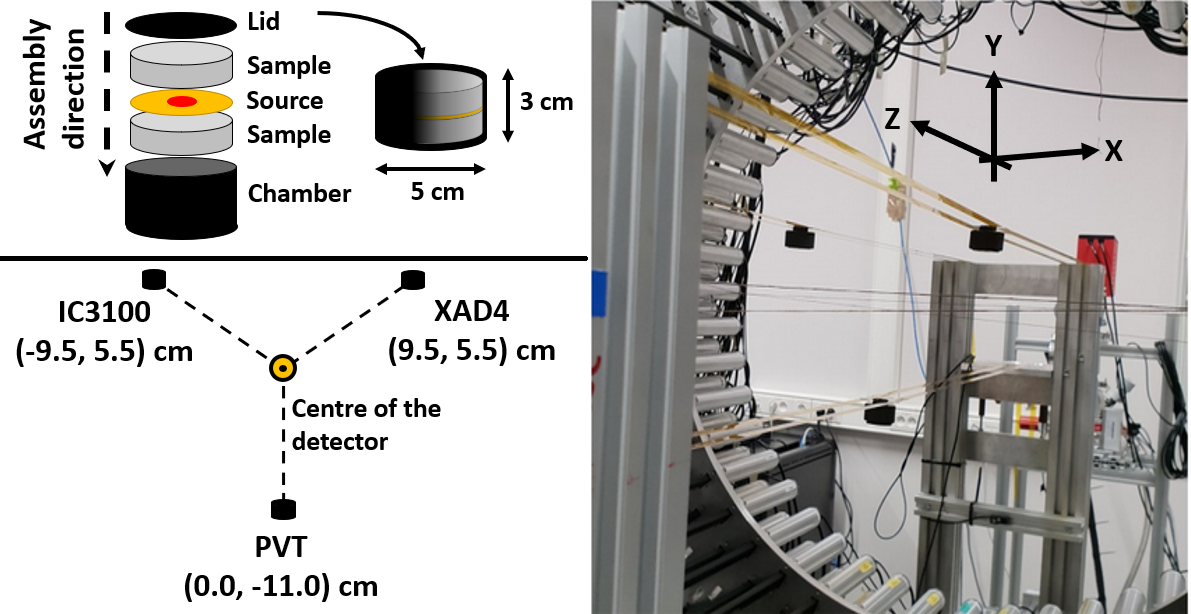}
    \caption{(left) Scheme of the measurement setup and plastic chambers used in the measurement with three materials - IC3100, XAD4 and PVT. In the brackets positions in XY plane are given. In the axial direction (Z) the samples were placed in the middle of the scintillators in order to obtain highest detection efficiency. Each chamber was equidistant from the centre of the detector, hanged on Kapton foil scaffolding. (right) Photo of the setup inserted into the detector. Chambers were placed on a Kapton scaffolding.}
    \label{fig:Measurement_setup}
\end{figure}

\noindent Electric signals are sampled at four voltage thresholds and converted to the time stamps by front-end electronics \cite{JINST-FPGA-2017}. The data are acquired by a trigger-less system \cite{IEEE-FPGA-2018} and analyzed using dedicated \textit{Framework} \cite{Science-Morpho-2024, Nature-Discrete-2024}. In the J-PET detector, time of the signal is estimated as the time on the lowest threshold with resolution of 350 ps (FWHM), which corresponds to the spatial resolution of 2 cm (standard deviation) \cite{EPJC-OrthoDecayoPS-2016}, when position of positron-electron annihilation is reconstructed based on the time. Next generation J-PET prototypes and development of new materials for better timing \cite{EPJC-OrthoDecayoPS-2020} allow for further improvement of spatial resolutions \cite{EJNMMI-Morpho-2020, PETCLin-TBJPET-2020, EJNMMI-TBJPET-2020}. The new J-PET prototypes should be also characterized by a higher sensitivity, therefore allowing an additional reduction of the measurement time \cite{EJNMMI-Morpho-2020, PETCLin-TBJPET-2020}.

\begin{figure}[ht!] 
    \centering
    \includegraphics[width=0.33\textwidth]{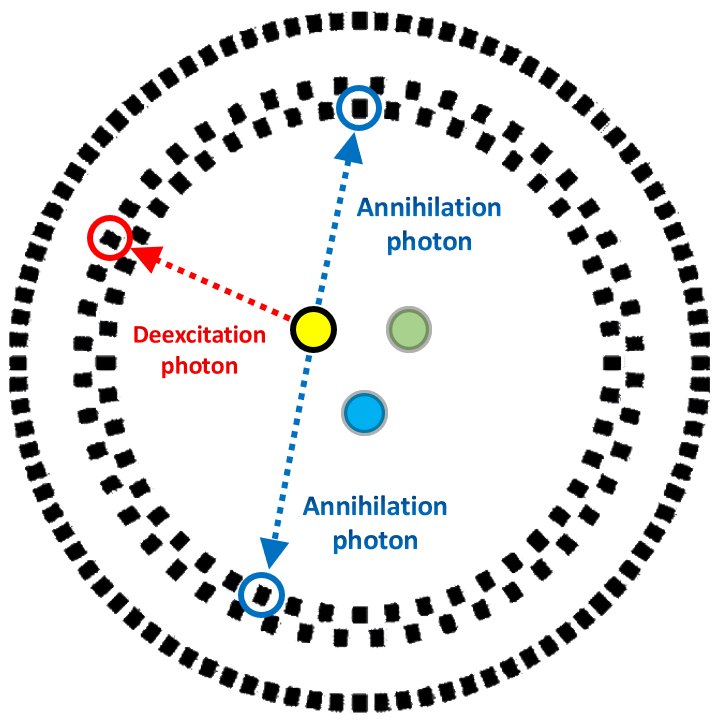}
    \caption{Measurement setup consisted of three samples (circles), each marked with different colour - IC3100 with yellow, XAD4 with green and PVT with blue, that were inserted to the J-PET detector (black rectangles). Example of the event used for the imaging and lifetime estimation coming from the sample marked as yellow - two annihilation photons (blue dashed arrows) and one deexcitation photon (red dashed arrow) registered in the J-PET detector.}
    \label{fig:Event}
\end{figure}

\noindent A measurement with three materials (XAD4 powder \cite{XAD4}, IC3100 powder \cite{IC3100} and PVT - scintillator \cite{PVT}) that differ in mean lifetime of ortho-positronium \cite{ActaB-3gammaFraction-2016} was conducted in order to check the ability of the J-PET detector to spatially characterize mean lifetime of positronium during a single measurement. The experimental setup is shown in Fig.~\ref{fig:Measurement_setup}. Three $^{22}$ Na radioactive sources (activity 0.24, 0.39 and 0.35 MBq respectively) in thin Kapton foil were used as positron emitters. $^{22}$Na nucleus during $\beta ^+$ decay emits a positron with an energy of about 545 keV and transforms into the excited state of $^{22}$Ne, which after about 4 ps emits a deexcitation photon with an energy of 1275 keV \cite{BAMS-DetEff-2023}. Each source was surrounded by a different material and inserted into a plastic cylindrical chamber shown schematically in Fig.~\ref{fig:Measurement_setup}. To keep the same detection efficiency, each chamber was placed at the same distance from the center of the detector.
\begin{figure}[H]
    \centering
    \includegraphics[width=0.43\textwidth]{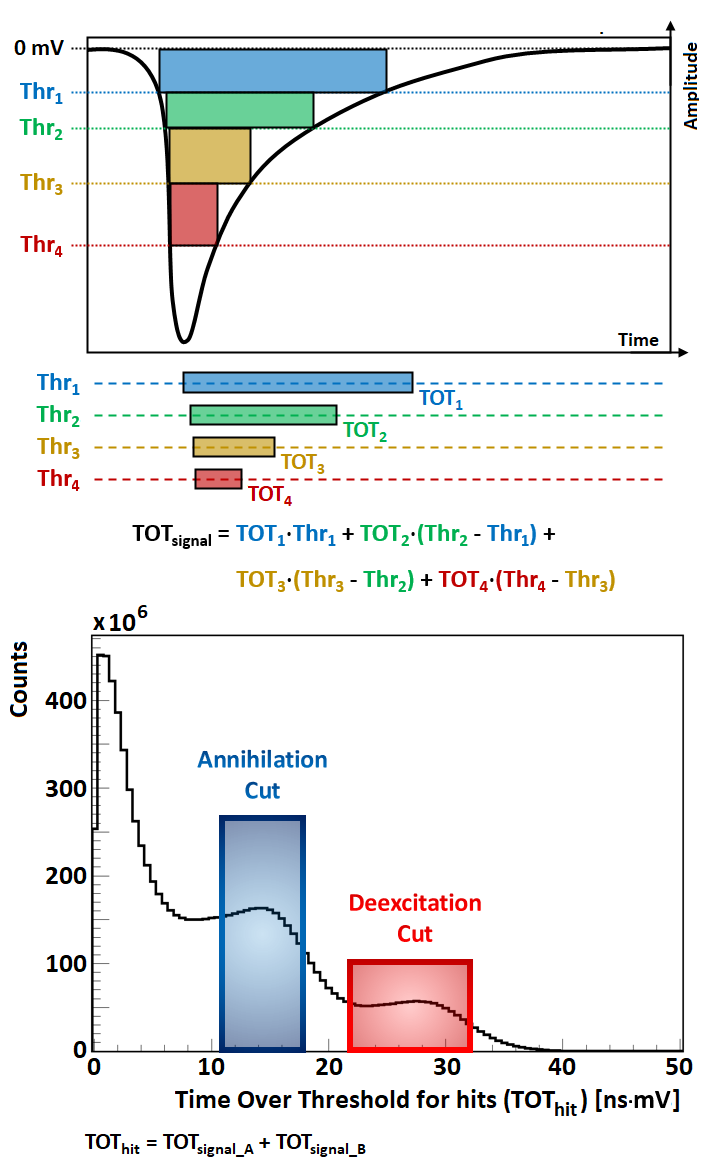}
    \caption{(top) Pictorial scheme of the Time Over Threshold (TOT) definition for the \mbox{J-PET} detector. TOT of the signal is a weighted sum of the TOTs for individual thresholds, with weights depending on the threshold value (Thr). (bottom) Hit TOT distribution. A single strip is connected to two photomultipliers (A and B) and the hit TOT is defined as the sum of signal TOTs for both ends of the strip. Selection criteria for annihilation and deexcitation photons are marked with blue and red, respectively.}
    \label{fig:TOT}
\end{figure}

\noindent Conditions (temperature 21$^{\circ}$C $\pm$ 0.5$^{\circ}$C, pressure 997.7 $\pm$ 2 hPa, humidity 58.9 $\pm$ 7.5 \%) were stable during the measurement.
Fig.~\ref{fig:Event} indicates a cross section of the J-PET detector with the superimposed arrows depicting photons from the signal event: annihilations (blue) and deexcitation (red). For each registered photon a time and position of interaction is calculated. Position and time of annihilation photons interaction are used to reconstruct the annihilation position \cite{ActaB-JPET-2017} while the time difference between emission of annihilation and deexcitation photons enables to calculate the positron lifetime for each registered event separately. Annihilation and deexcitation photons are disentangled based on the measurement of the time-over-threshold (TOT) of registered electric signals. Definition of TOT is presented graphically in Fig.~\ref{fig:TOT}. TOT scales with the photon energy as it was shown recently in the reference \cite{EJNMMI-TOT-2020}. 

\section*{IMAGING AND LIFETIME ANALYSIS}

The analysis of the data collected from the measurement was divided onto several steps: data selection, background reduction, annihilation position distribution reconstruction and PALS analysis for each voxel of the imaged objects. The resulting positron lifetime distribution was decomposed by a dedicated software PALS Avalanche \cite{ActaA-PAv-2017, ActaA-PAv-2020}.

\begin{figure}[H] 
    \centering
    \includegraphics[width=0.45\textwidth]{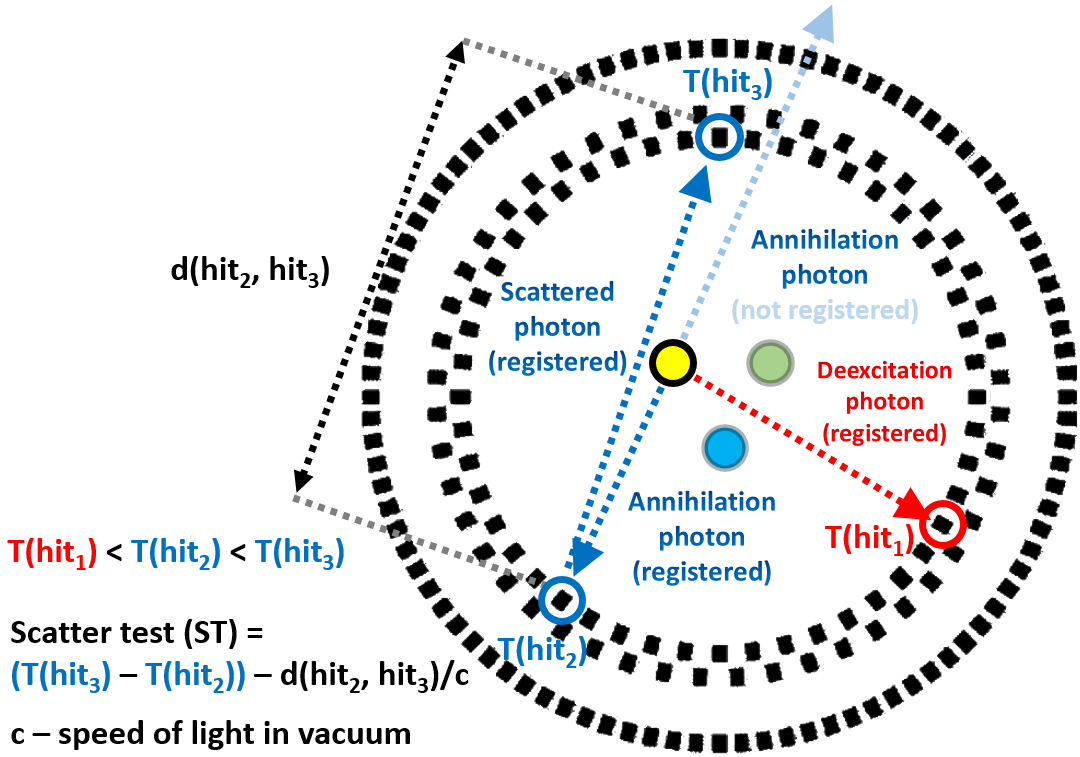}
    \caption{Exemplary scheme of an event with the scattering of the annihilation photon. One of the annihilation photon scatters in the detection module and its scattering is additionally registered in the second module. In addition, the second annihilation photon is not registered. Such scatterings can be rejected by using scatter test defined in the figure. }
    \label{fig:ExampleScatt}
\end{figure}

\subsection*{DATA SELECTION}
The first step was focused on the reconstruction of the signals, hits and events from the trigger-less data acquisition. A signal corresponds to eight timestamps measured at four threshold levels at the leading and trailing edges of the electric pulse from a single photomultiplier. The time of the signal is taken as the time on the lowest threshold. A pair of signals within 50 ns time window from photomultipliers connected to the same scintillator strip formed a hit. A hit is defined as a reconstructed interaction point characterized by the time, position and TOT. A set of hits in 200 ns time window formed an event. An example of the event that will allow the reconstruction of the annihilation position and positron lifetime estimation is schematically shown in Fig.~\ref{fig:Event}. After reconstructing all of the events, further analysis was conducted on the event-by-event basis. For each hit in the event, the energy-equivalent TOT value was calculated \cite{EJNMMI-TOT-2020}. Definition of the TOT of the signal can be seen in Fig.~\ref{fig:TOT}(top). Based on the TOT value, a hit was assigned to one of the two categories:

\begin{figure*}[ht!] 
    \centering
    \includegraphics[width=0.95\textwidth]{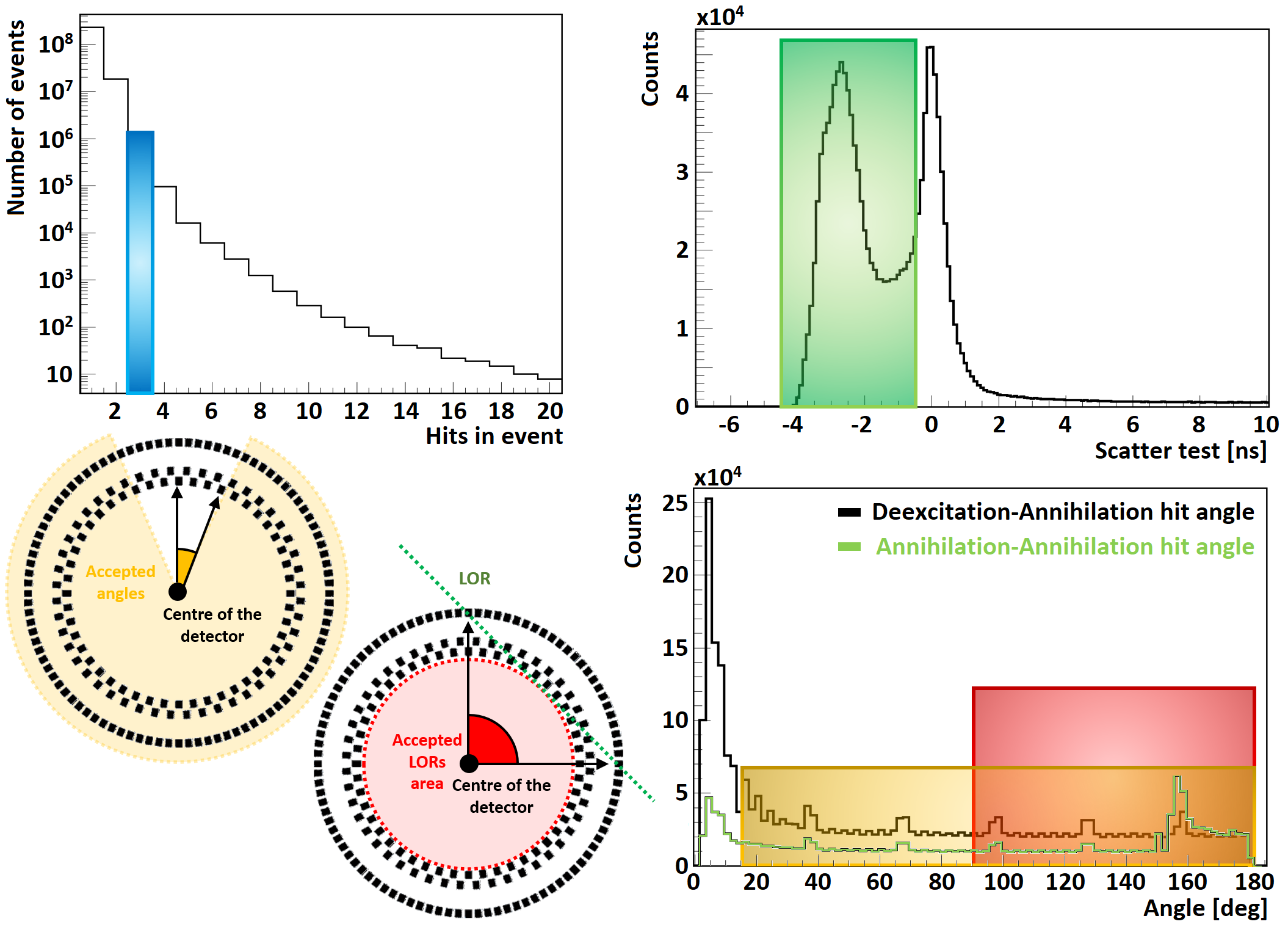}
    \caption{(left, top) Multiplicity distribution of hits in event. Blue rectangular indicates which event multiplicity was selected for further analysis. Tail coming from multiplicity 4 to 20 shows the influence of  multiple scatterings, the positron-electron decays on more photons, accidental coincidences, and cosmic rays. (right, top) Scatter test distribution for candidates for annihilation hits. One can see three structures, the first peak for negative values correspond to the true back-to-back pair with very small time differences, which exclude the possibility of scattering. The second peak around zero comes from the true scatterings in which time difference is comparable to the time a photon would need to travel between two scintillators. At last the long tail for positive values of the scatter test originates from the accidental coincidences and misidentification of the annihilation with the deexcitation hit or its scattering. Green area indicates range of acceptance of events. (left, bottom) Selected ranges of angles between any annihilation and deexcitation hits (orange) and between annihilation hits only (red) selected for further analysis. Angle between annihilation hits can be translated to the range of LORs, that are accepted. (right, bottom) Distribution of the angles between any two hits. The stepped structure of the distribution comes from the geometry of the J-PET detector and the enhancement for some bins comes from inclusion of two scintillators within one bin every 30$^{\circ}$, starting from 7.5$^{\circ}$ bin, which is the period of the first layer of scintillators (left, bottom). The peak near zero originates from the scatterings between neighbouring scintillators. Orange area indicates the acceptance range of the angle between deexcitation and annihilation hits, to reduce scatterings between neighbouring scintillators. Red area indicates the acceptance range of the angle between hits classified as annihilation candidates to ensure that annihilation took place inside the detector. }
    \label{fig:SelectionCriteria}
\end{figure*}

\[
    \text{Annihilation hits where } 11 \leq \text{ TOT}_{\text{hit}} \leq 18,
\]
\[
    \text{Deexcitation hits where } 22 \leq \text{ TOT}_{\text{hit}} \leq 32.
\]

\noindent To establish a reference time (emission of a deexcitation photon) and estimate the lifetime of a positron, at least two annihilation and one additional deexcitation hit must be a part of the same event. The reconstructed position is assigned to an event (including the deexcitation hit from the event) based on the positions and times of the annihilation hits from that event. The simplest and most common case of an event with these characteristics was selected, which consisted of two annihilation hits along with one deexcitation hit. The distribution of hit TOT values with selection ranges superimposed for the two categories is
shown in Fig.~\ref{fig:TOT}(bottom).

\subsection*{BACKGROUND REDUCTION}

To reduce the impact of various background sources that can cause artifacts the final images, such as secondary scattering of photons in the detector (shown schematically in Fig.~\ref{fig:ExampleScatt}), cosmic rays, random coincidences and electronic noises, a set of selection criteria has been introduced. The selection criteria set consisted of:

\begin{enumerate}
    \item Multiplicity cut - to avoid ambiguity of proper selection of hits in event, number of hits in event (multiplicity) should be exactly three (where two of them should be categorized as annihilation, and the third one as deexcitation) as shown in Fig.~\ref{fig:SelectionCriteria} (left, top);
    \item Scatter test - to reduce the misidentification of secondary scattered photon as the annihilation hit (Fig.~\ref{fig:ExampleScatt}), a scatter test (ST) was introduced as:
    \[ \text{ST} = \Delta\text{t}\left( hit_1, hit_2\right) - \frac{\text{d}\left( hit_1, hit_2\right)}{\text{c}} \]
    where $\Delta$t is a time difference between registration of two hits, d is a distance between two hit positions, and c is the speed of light in vacuum. In case of the misidentification due to the scattering the value of ST is equal to zero. To avoid disturbing the positron lifetime distribution, which is based on a reference time of deexcitation hit, the scatter test was performed only for the hits that were assigned as annihilation hits. Cut-off for the test was selected as -0.5~ns $\left( \text{ST} \leq -0.5 \text{ ns}\right)$ as shown in Fig.~\ref{fig:SelectionCriteria} (top right);
    \item Angular cuts - to additionally reduce scatterings between neighbouring scintillators, that could survive the previous condition, the minimal angle between any hits in event should be greater than $15^{\circ}$. Moreover, in order to further reduce contribution of scatterings, the angle between annihilation hits, measured from the center, was required to be greater than 90$^{\circ}$. Angle cuts and the schemes of the geometries rejected by the used condition is shown in Fig.~\ref{fig:SelectionCriteria} (bottom).
\end{enumerate}

\begin{figure}[ht!] 
    \centering
    \includegraphics[width=0.45\textwidth]{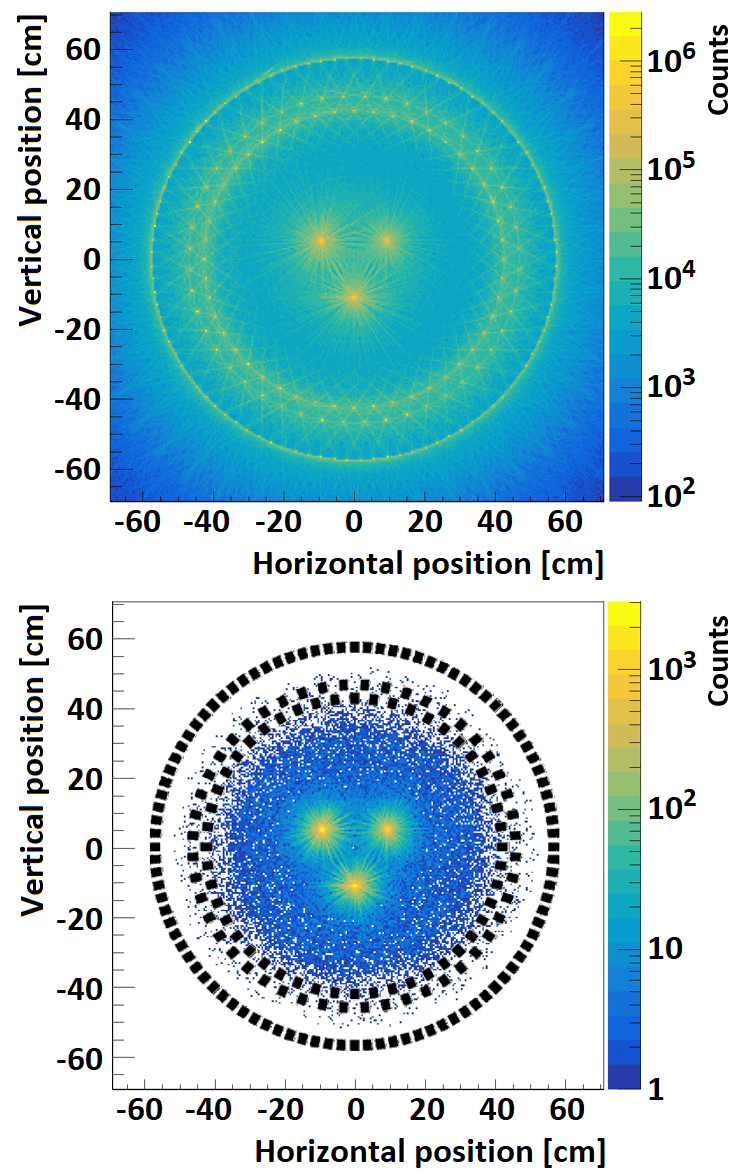}
    \caption{(top) Distribution of all of the annihilation positions before applying selection criteria. The green circles covering the scintillator positions are from events in which both classified annihilation hits came from the scattering of a single photon, while other annihilation photon was not registered.
   % Green circle with the greatest radius (57.5 cm) is coming from events in which the annihilation photon was detected and scattered in the 3$^{rd}$ layer of scintillators and then scattering was detected in other scintillator and classified as the second annihilation hit. Green ring that has the second greatest radius (approximately 44 cm) are the scatterings that happen in the first and the second layer of scintillators, and then were detected in other scintillator and classified as annihilation hit. 
    (bottom) Distribution of the annihilation positions after applying selection criteria. To guide an eye black rectangles indicate positions of the scintillators. There are still some scatterings and accidental coincidences ($\approx 0.7 \% $)
    % (182+186+181+212) * 2 / 216579 = 0.7 % of those between black rectangles
    % 216579 - 203700 + 25200 / 216579 = 15.3 % of all BG besides sources
    that have survived the selection criteria used, visible between the black rectangles.}
    \label{fig:SuvImages}
\end{figure}

\subsection*{POSITION RECONSTRUCTION}

At the last stage of the analysis, image is created from the events that have fulfilled the entire selection criteria. From selected events, the position of annihilation is estimated based on the position and time of registration of the annihilation hits:
\begin{equation}
\frac{\text{P}\left( hit_1\right) + \text{P}\left( hit_2\right)}{2} + \left( \text{T} \left( hit_1\right) - \text{T} \left( hit_2\right) \right)\times \text{ c } \times \widehat{P} _{1 \rightarrow 2} ,
\end{equation}

\noindent where P is a position of the hit, T is the time of the hit, c is the speed of light in vacuum and $\widehat{P} _{1 \rightarrow 2}$ is a versor directed from the P($hit_1$) to P($hit_2$). A projection onto XY plane of the reconstructed annihilation points, forming an image of the $\beta ^+$ activity distribution, before and after applying the selection criteria is shown in Fig.~\ref{fig:SuvImages}. In the image one can clearly see the positions of each sample, with no clear distinction between different materials. The reconstructed annihilation rate should be proportional to the activity of the $\beta ^+$ source. Which is confirmed by calculating the ratio of events in the image to the source activity (second row in Tab.~\ref{tab:PALS}).

\section*{POSITRONIUM LIFETIME IMAGING}

\subsection*{POSITRON LIFETIME DISTRIBUTION}

\begin{figure}[ht!] 
    \centering
    \includegraphics[width=0.45\textwidth]{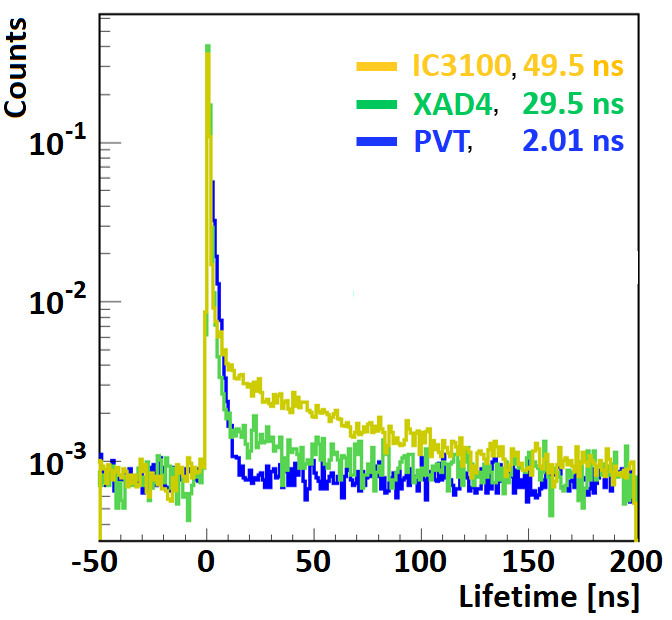}
    \caption{Positron lifetime distribution for each material used in the measurement. Distributions obtained from hits positioned inside three regions of interest are marked with appropriate colors. Distributions are normalized such that the accidental coincidences background is on the same level. Fitting each distribution resulted in obtaining mean o-Ps lifetime (longest component) for each sample - 49.5 ns for IC3100 (yellow), 29.6 ns for XAD4 (green) and 2.01 ns for PVT (blue). }
    \label{fig:PALS}
\end{figure}

At each location in the activity image (voxel), one can determine the positron annihilation lifetime spectrum based on the reference time coming from the deexcitation photon. Positron lifetime for a single event can be estimated as:
\begin{equation}
\frac{ \left| \text{T} \left( hit^{Anni}_1\right) + \text{T} \left( hit^{Anni}_2\right) \right|}{2} - \text{T} \left( hit^{Deex}\right),
\end{equation}

\noindent where T is the time of a hit, $hit^{Anni}_1$ and $hit^{Anni}_2$ are the annihilation hits selected in a given event, and $hit^{Deex}$ is the deexcitation hit in this event. The lifetime spectrum of positronium may be determined for each voxel of the annihilation rate image. As an example, Fig.~\ref{fig:PALS}  shows exemplary positron lifetime distribution for voxels inside the circles with radius of 2.5 cm around the center of the samples which contain most of the activity inside. The spectrum comprises contributions from annihilation of p-Ps, direct electron-positron annihilation, annihilation in the material surrounding the source, background from accidental coincidences, and the o-Ps components. The contribution from the long-lived o-Ps is clearly visible, especially in the spectrum of IC3100 material (Fig.~\ref{fig:PALS}) for which the spectrum approaches the accidental background level only at about 200 ns. During the fitting, the number of o-Ps components was set to two for XAD4 and IC3100 and one for PVT, which was consistent with the literature results and gave good fits measured by the adjusted R$^2$ value.
% 0.9885 -> 0.9889 XAD4 ; ; 0.9780 -> 0.9789 IC3100
%\begin{figure}[h!] 
%    \centering
%    \includegraphics[width=0.45\textwidth]{Porous_phantom_fits_IC3100.png}
%    \caption{(top) Fit to the positron lifetime distribution collected from the annihilation positions near the position of the IC3100 sample, shown in full and restricted range. (bottom) Components of each of the positron-electron decay channel, from the fit, drawn seprately to the positron lifetime spectrum for IC3100.}
%    \label{fig:IC3100Fit}
%\end{figure}

\begin{figure}[ht!] 
    \centering
    \includegraphics[width=0.45\textwidth]{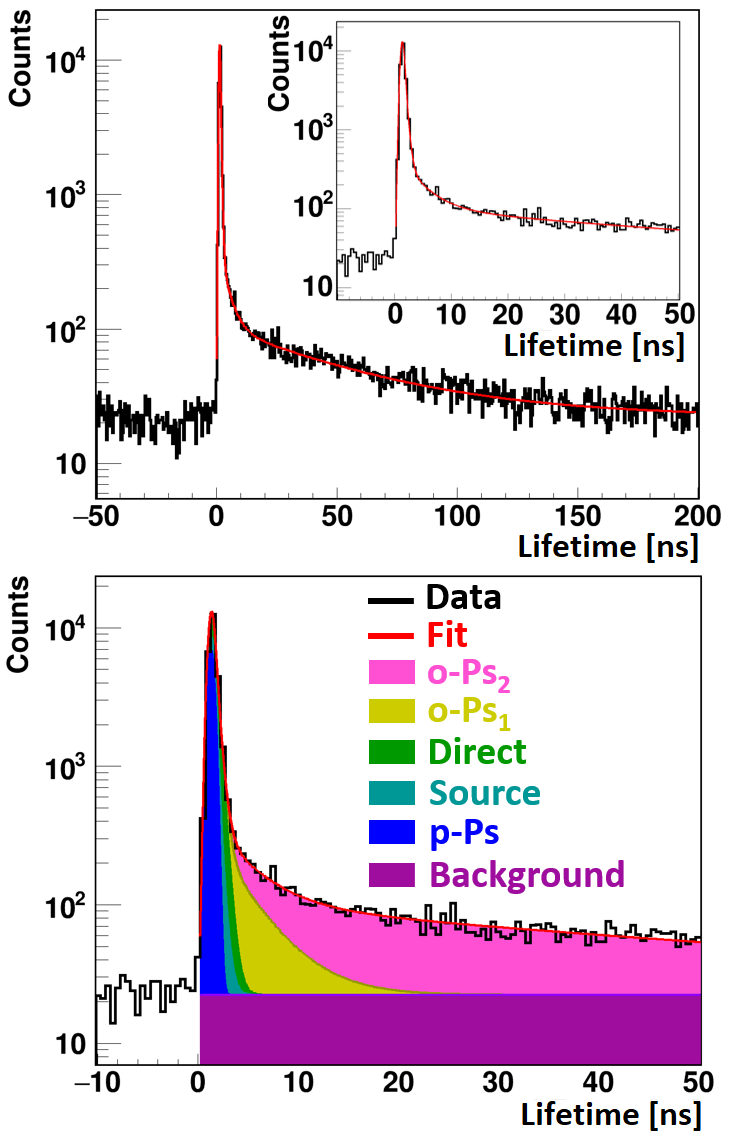}
    \caption{(top) Positron lifetime spectrum determined for the IC3100 sample. Superimposed red curve indicates the result of the fit. (bottom) Positron lifetime spectrum (as above). Superimposed histograms indicate in a cumulative way, contributions to the spectrum from o-Ps annihilations, direct annihilation, annihilation in the source, p-Ps annihilation and the background.}
    \label{fig:IC3100Fit}
\end{figure}

\noindent Results from decomposition into different positron lifetime components by fitting each positron lifetime distribution from Fig.~\ref{fig:PALS} are shown in Table~\ref{tab:PALS} and exemplary decomposed spectrum for the IC3100 is shown in Fig.~\ref{fig:IC3100Fit}. Fitting was performed by PALS Avalanche software \cite{ActaA-PAv-2017, ActaA-PAv-2020}. Contribution from the annihilations in Kapton foil that contained the $^{22}$Na source, was fixed to the known values of 10$\%$ intensity and 0.374 ns mean lifetime \cite{ActaB-3gammaFraction-2016}. As the most sensitive structural probe is a o-Ps mean lifetime distribution, to stabilize fitting procedure, mean lifetime of the p-Ps was fixed to the theoretical value in vacuum amounting to 0.125 ns \cite{PhysRevLett-pPsDecayRate-1994}.

\begin{table}[ht!]
    \caption{Results from fitting positron annihilation lifetime spectra for three materials used for the measurement - IC3100, XAD4, PVT. Contribution of the annihilation in the Kapton foil was excluded from the total intensity of the positron annihilation.} 
    \label{tab:PALS}
    \centering
    \begin{tabular}{|c|c|c|c|}
        \hline
       \textbf{ Parameter name} & \textbf{IC3100} & \textbf{XAD4} & \textbf{PVT}\\
        \hline
        \hline
        Total number of counts & 53542 & 35313 & 52408 \\
        \hline
        Activity of the source [MBq] & 0.393 & 0.238 & 0.345\\
        \hline
        p-Ps mean lifetime & 0.125 & 0.125 & 0.125 \\ 
        (fixed parameter) [ns] & & & \\
        \hline
      %  p-Ps intensity [$\%$] & 35.98 (53) & 23.99 (93) & 29.52 (52)\\
        p-Ps intensity [$\%$] & 21.98 (61) & 13.41 (98) & 22.47 (55)\\
        \hline
      %  direct annihilation & 0.435 (09) & 0.404 (07) & 0.452 (15)\\
        direct annihilation & 0.385 (06) & 0.383 (06) & 0.408 (11)\\
        mean lifetime [ns] & & &\\
        \hline
      %  direct annihilation intensity [$\%$] & 31.13 (28) & 55.89 (28) & 25.76 (53) \\
        direct annihilation intensity [$\%$] & 45.18 (28) & 65.44 (28) & 31.91 (51) \\
        \hline
      %  o-Ps mean lifetime 1 & 3.64 (23) & 1.65 (05) & 2.02 (02)\\
        o-Ps mean lifetime 1 & 3.39 (21) & 1.64 (05) & 2.01 (02)\\
        component [ns] & & &\\
        \hline
       % o-Ps intensity 1 & 6.65 (07) & 15.29 (02) & 44.71 (52) \\
        o-Ps intensity 1 & 6.97 (07) & 15.89 (01) & 45.61 (51) \\
        component [$\%$] & & &\\
        \hline
       % o-Ps mean lifetime 2 & 51.4 (1.2) & 25.8 (2.6) & -\\
        o-Ps mean lifetime 2 & 49.5 (1.2) & 29.5 (2.8) & -\\
        component [ns] & & & \\
        \hline
      %  o-Ps intensity 2 & 26.24 (35) & 4.82 (29) & -\\
        o-Ps intensity 2 & 25.86 (34) & 5.25 (29) & -\\
        component [$\%$] & & &\\
        \hline
      %  Reduced R Squared & 0.9961 & 0.9958 & 0.9976\\
        Reduced Chi Squared & 0.9682 & 1.0778 & 1.1527\\
        \hline
    \end{tabular}
\end{table}

\noindent %However, the main challenge for this article is to demonstrate that the J-PET detector is able to determine mean positronium lifetime in different materials simultaneously and therefore estimate nanoporosity in each voxel of the investigated object. 
Results from the Table~\ref{tab:PALS} indicate that, the differences of the o-Ps mean lifetime in the investigated materials are in the order of tens of nanoseconds. It is worth to notice, that the measurement was conducted in the atmospheric pressure, so due to the conversion processes, the obtained results differ from literature values (131.9 ns for IC3100, 90.8 ns for XAD4, 2.05 ns for PVT)\cite{PVT, ActaB-3gammaFraction-2016}, that were obtained from measurements in vacuum.

\subsection*{LIFETIME IMAGING AND NANOPOROSITY MAP}

\begin{figure}[ht!] 
    \centering
    \includegraphics[width=0.48\textwidth]{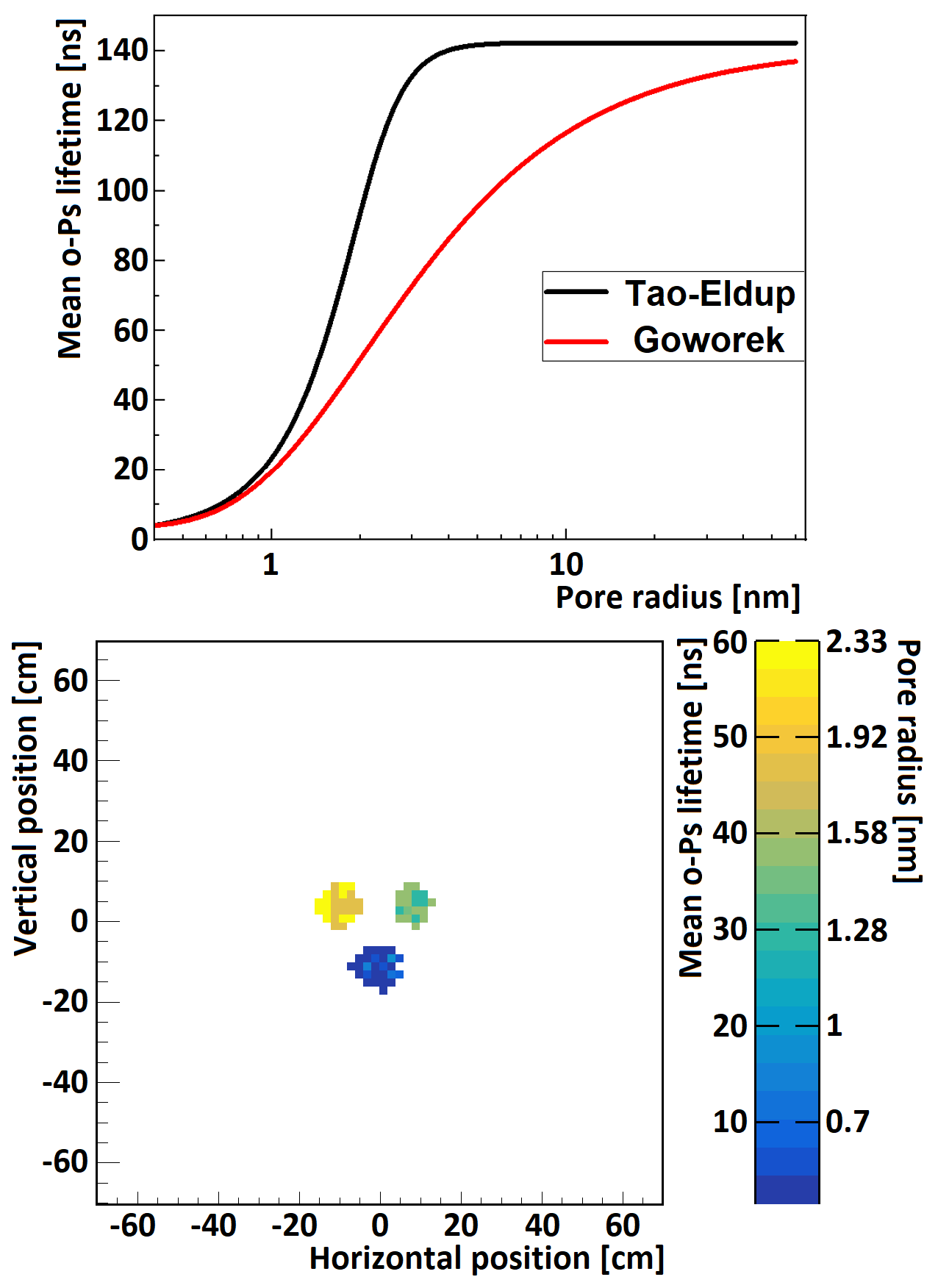}
    \caption{
    (top) Mean o-Ps lifetime as a function of the mean pore radius, following Tao-Eldrup and Goworek model \cite{ChemPhys-LFvsSize-1981, ChemPhysLett-LFvsSize-1997}. (bottom) Positron lifetime map constructed from the mean ortho-positronium lifetime estimates for voxels where the total number of counts were at least 5$\%$ of the maximum total number of counts for a single voxel (67 voxels visible). There is a visible difference between three areas which are representing materials that differ in mean positron lifetime, according to Table~\ref{tab:PALS} (top left - IC3100, top right - XAD4, bottom - PVT). Using Goworek model mean o-Ps lifetime scale was recalculated to the pores radius. }
    \label{fig:LFimage}
\end{figure}

For the next step, an image voxel was established to the size of 1 $\times$ 1 $\times$ 6 cm$^3$, from which the distribution of positron lifetimes was collected from. The voxel size in the axial direction was larger due to the poorer resolution in this dimension (standard deviation 0.72 cm for the horizontal and vertical directions and 3 cm for the axial direction) \cite{ActaB-JPET-2017}. This made it possible to obtain about several thousand counts per voxel near the sources. Resulting positronium lifetime image for defined voxels is shown in Fig.~\ref{fig:LFimage}.
The distribution of pore size in the imaged materials can be estimated based on the mean lifetime of the o-Ps components from the Tao-Eldrup model \cite{ChemPhys-LFvsSize-1972, ChemPhys-LFvsSize-1981} and its modifications \cite{PALS_porosity1, PALS_porosity2, ChemPhysLett-LFvsSize-1997}. The relation between the mean o-Ps lifetime and the mean radius of the free voids is shown in the upper panel of Fig.~\ref{fig:LFimage}. It is worth emphasizing that the impact of various processes influencing the shortening of the mean o-Ps lifetime, such as pick-off and ortho-para conversion, was not distinguished.
%In the final step of analysis, pore size distribution for each voxel was estimated, showing capabilities of the J-PET detector to scan voxels in the nanoporosity domain. 
%As it was shown in Fig.~\ref{fig:PALS} the J-PET detector is able to obtain positron lifetime distribution from a given detection area and to successfully decompose it onto different components. In the next step, a procedure will be described to estimate the mean positronium lifetime estimate in each voxel. 

\noindent 
To increase the contrast only the highest intensity voxels are shown, where the total number of counts is at least 5$\%$ of the maximum total number of counts for a single voxel. It is clearly visible that there is a difference in the pore size between three regions corresponding to different materials used in the measurement.

% \begin{figure}[h!] 
%     \centering
%     \includegraphics[width=0.35\textwidth]{Propozycja_Rysunku.png}
%     \caption{
%     Reconstructed positron lifetime map after XX iteration, shown for voxels where the total number of counts were at least 5$\%$ of the maximum total number of counts for a single voxel. There is a visible difference between three areas which are representing materials that differ in mean positron lifetime, according to Table~\ref{tab:PALS} (top left - IC3100, top right - XAD4, bottom - PVT). Algorithm allowed to reach diameters of hot areas 1.7 cm for XY and 4.2 cm in Z. }
%     \label{fig:MLEMoPsLFimage}
% \end{figure}

%\begin{figure}[h!] 
%    \centering
%    \includegraphics[width=0.38\textwidth]{VerticalPI3src_2.png}
%    \caption{
%    Transverse (top, $z = 0$~cm) and coronal (bottom, $y=5.5$~cm) cross-sections of the positron lifetime map, reconstructed after the $6$-th MLEM iteration for the voxels where the activity exceeded 5$\%$ of the maximum (761 in total). }
%    \label{fig:MLEMoPsLFimage}
%\end{figure}

\section*{CONCLUSIONS}

Combining information from the positronium decay - position of annihilation and lifetime of positronium, gives the possibility of a more complex analysis of materials. The simultaneous spatial and structural characteristics of materials can be used in the study of structural changes in sample, diffusion inside the sample and many others. 

\noindent The measurement with three materials (IC3100, XAD4, PVT) differing in the mean positronium lifetime was carried out using the J-PET detector. Dedicated data selection was used to reduce various background sources such as scatterings, cosmic rays and electronic noise. Analysis of the positron annihilation lifetime distribution sampled in three different areas showed the J-PET detector's ability to accurately estimate the mean positron lifetime of different components (p-Ps, o-Ps, direct annihilation) in each area separately.

%\noindent As the final step of the analysis mean positron lifetime image was created for each voxel from the annihilation rate image. Lifetime image allowed spatial differentiation of different materials based on the positron mean lifetime estimate value in each voxel. The use of a positronium imaging reconstruction algorithm dedicated to the J-PET detector \cite{Shopa2022} allowed for a significant improvement in image resolution to approximately 1.7 cm in the horizontal and vertical direction and approximately 4.2 cm in the axial direction. 

\noindent This highlights J-PET as the first multiphoton PET system with the ability to obtain a porosity image based on positron annihilation. Further development of detectors in the context of time resolution \cite{Lecoq2020} as well as significant improvement in sensitivity based on Total-Body projects \cite{TotalBodyJPET} may also lead to an improvement in the quality of positronium images, including even millimeter resolution. 
%Further studies would be focused on the resolution determination of the lifetime imaging and optimizing of the analysis and measurement procedures.

\noindent Extending the capabilities of the PET allowing for much earlier diagnosis can improve the effectiveness of cancer therapy. It is also worth noting, that positron annihilation may become a sensitive probe also in fundamental research, especially in the test of discrete symmetry violation \cite{PhysRev-Csymmetry-1967, PhysRevA-Csymmetry-1996, PhysRevA-Csymmetry-2002, PhysRev-CPsymmetry-2010, PhysRev-CPsymmetry-1991, PhysRevA-CPTsymmetry-1988, PhysRev-CPTsymmetry-2003, ActaB-QED-2019, Symmetry-symmetry-2020, Nature-Discrete-2021, Nature-Entaglement-2024}, quantum entanglement \cite{Nature-Discrete-2021, ActaB-Entanglement-2017, Scientific-Entanglement-2017, Scientific-Entanglement-2019} and forbidden decays \cite{ActaBSupp-Invisible-2024, PhysRev-Invisible-2020}.

\section*{AUTHOR CONTRIBUTION}

The experiment was conducted using the Jagiellonian Positron Emission Tomograph (J-PET). The J-
PET scanner, the techniques of the experiment and this study was conceived by P.M. The preparation of the experimental setup and samples was done by K.D. The data analysis was conducted by K.D. Signal selection criteria were developed by P.M. and K.D. and applied by K.D. Authors: K.D., E.B., N.C., C.C., E.C., M.D., M.G., B.J., K.Kacprzak, Ł.K., G.K., T.K., K. Kubat, D.K., E.L., F.L., J.M-S., S.N., P.P., S.P., E.PdR., M.R.m S.S., M.S. E.Ł.S., K.T., P.T. and P.M participated in the construction, commissioning, and operation of the experimental setup, as well as in the data-taking campaign and data interpretation. K.D. under the supervision of P.M. monitored the whole data taking campaign. M.G. and B.J. designed and constructed the positronium production chamber. G.K. and S.N. optimized the working parameters of the detector. K.D. and K. Kacprzak, took part in developing the J-PET analysis and simulation framework. K.D., M.S., K.K. and E.PdR. performed timing calibration of the detector. E.C. developed and operated short- and long-term data archiving systems and the computer center of J-PET. S.S. established relation between energy loss and TOT and dependence of detection efficiency on energy deposition. P.M. managed the whole project and secured the main financing. The results were interpreted by K.D. and P.M.. The manuscript was prepared by K.D. and P.M. and was then edited and approved by all authors.

\section*{ACKNOWLEDGMENT}

The authors would like to acknowledge the technical support from A. Heczko, M. Kajetanowicz, and W. Migdał. The authors are also grateful to the J-PET members for their editorial remarks on the earlier version of the manuscript. We acknowledge the support provided by the National Science Centre of Poland through grants no. 2021/42/A/ST2/00423 (P.M.), 2021/43/B/ST2/02150 (P.M.), 2022/47/I/NZ7/03112 (E.Ł.S) and 2021/41/N/ST2/03950 (K.D.); the Ministry of Education and Science through grants no. SPUB/SP/490528/2021 (P.M.), IAL/SP/596235/2023 (P.M.); as well as the SciMat and qLife Priority Research Area budgets under the programme Excellence Initiative - Research University at the Jagiellonian University (P.M.).

\section*{REFERENCES}

\bibliographystyle{plain}

\end{document}